\documentclass[useAMS,usenatbib]{mn2e}

\usepackage{xspace}
\usepackage{graphicx}
\usepackage{tabularx}
\newcommand{\ignore}[1]{}
\providecommand{\ao}{}
\renewcommand{\ao}{adaptive optics (AO)\renewcommand{\ao}{AO\xspace}\renewcommand{\Ao}{AO\xspace}\xspace}
\newcommand{\Ao}{Adaptive optics (AO)\renewcommand{\ao}{AO\xspace}\renewcommand{\Ao}{AO\xspace}\xspace}
\newcommand{\wfs}{wavefront sensor (WFS)\renewcommand{\wfs}{WFS\xspace}\renewcommand{\wfss}{WFSs\xspace}\xspace}
\newcommand{\wfss}{wavefront sensors (WFSs)\renewcommand{\wfs}{WFS\xspace}\renewcommand{\wfss}{WFSs\xspace}\xspace}
\newcommand{\shwfs}{Shack-Hartmann \wfs (SHWFS)\renewcommand{\shwfs}{SHWFS\xspace}\xspace}
\newcommand{\dm}{deformable mirror (DM)\renewcommand{\dm}{DM\xspace}\renewcommand{\dms}{DMs\xspace}\renewcommand{\Dms}{DMs\xspace}\renewcommand{\Dm}{DM\xspace}\xspace}
\newcommand{\dms}{deformable mirrors (DMs)\renewcommand{\dm}{DM\xspace}\renewcommand{\dms}{DMs\xspace}\renewcommand{\Dms}{DMs\xspace}\renewcommand{\Dm}{DM\xspace}\xspace}
\newcommand{\Dms}{Deformable mirrors (DMs)\renewcommand{\dm}{DM\xspace}\renewcommand{\dms}{DMs\xspace}\renewcommand{\Dms}{DMs\xspace}\renewcommand{\Dm}{DM\xspace}\xspace}
\newcommand{\Dm}{Deformable mirror (DM)\renewcommand{\dm}{DM\xspace}\renewcommand{\dms}{DMs\xspace}\renewcommand{\Dms}{DMs\xspace}\renewcommand{\Dm}{DM\xspace}\xspace}

\newcommand{\shs}{Shack-Hartmann sensor (SHS)\renewcommand{\shs}{SHS\xspace}\renewcommand{\shss}{SHSs\xspace}\xspace}
\newcommand{\shss}{Shack-Hartmann sensors (SHSs)\renewcommand{\shs}{SHS\xspace}\renewcommand{\shss}{SHSs\xspace}\xspace}
\newcommand{\lgs}{laser guide star (LGS)\renewcommand{\lgs}{LGS\xspace}\renewcommand{\lgss}{LGSs\xspace}\xspace}
\newcommand{\lgss}{laser guide stars (LGSs)\renewcommand{\lgs}{LGS\xspace}\renewcommand{\lgss}{LGSs\xspace}\xspace}
\newcommand{\ngs}{natural guide star (NGS)\renewcommand{\ngs}{NGS\xspace}\renewcommand{\ngss}{NGSs\xspace}\xspace}
\newcommand{\ngss}{natural guide stars (NGSs)\renewcommand{\ngs}{NGS\xspace}\renewcommand{\ngss}{NGSs\xspace}\xspace}
\newcommand{\mems}{Micro-Electro-Mechanical Systems (MEMS)\renewcommand{\mems}{MEMS\xspace}\xspace}
\newcommand{\snr}{signal to noise ratio (SNR)\renewcommand{\snr}{SNR\xspace}\xspace}
\newcommand{\moao}{multi-object \ao (MOAO)\renewcommand{\moao}{MOAO\xspace}\xspace}
\newcommand{\mcao}{multi-conjugate adaptive optics (MCAO)\renewcommand{\mcao}{MCAO\xspace}\xspace}
\newcommand{\ltao}{laser tomographic adaptive optics
  (LTAO)\renewcommand{\ltao}{LTAO\xspace}\xspace}
\newcommand{\glao}{ground layer adaptive optics (GLAO)\renewcommand{\glao}{GLAO\xspace}\xspace}
\newcommand{\cpu}{central processing unit (CPU)\renewcommand{\cpu}{CPU\xspace}\renewcommand{\cpus}{CPUs\xspace}\xspace}
\newcommand{\cpus}{central processing units (CPUs)\renewcommand{\cpu}{CPU\xspace}\renewcommand{\cpus}{CPUs\xspace}\xspace}
\newcommand{\psf}{point spread function (PSF)\renewcommand{\psf}{PSF\xspace}\renewcommand{\psfs}{PSFs\xspace}\xspace}
\newcommand{\psfs}{point spread functions (PSFs)\renewcommand{\psf}{PSF\xspace}\renewcommand{\psfs}{PSFs\xspace}\xspace}
\newcommand{\fpga}{field programmable gate array (FPGA)\renewcommand{\fpga}{FPGA\xspace}\renewcommand{\fpgas}{FPGAs\xspace}\xspace}
\newcommand{\fpgas}{field programmable gate arrays (FPGAs)\renewcommand{\fpga}{FPGA\xspace}\renewcommand{\fpgas}{FPGAs\xspace}\xspace}
\newcommand{\sor}{successive over-relaxation (SOR)\renewcommand{\sor}{SOR\xspace}\xspace}
\newcommand{\fdpcg}{Fourier domain pre-conditioned gradient (FDPCG)\renewcommand{\fdpcg}{FDPCG\xspace}\xspace}
\newcommand{\map}{maximum a-posteriori (MAP)\renewcommand{\map}{MAP\xspace}\xspace}
\newcommand{\elt}{Extremely Large Telescope (ELT)\renewcommand{\elt}{ELT\xspace}\renewcommand{\elts}{ELTs\xspace}\xspace}
\newcommand{\elts}{Extremely Large Telescopes (ELTs)\renewcommand{\elt}{ELT\xspace}\renewcommand{\elts}{ELTs\xspace}\xspace}
\newcommand{\dugall}{Durham University generalised adaptive optics laser laboratory (DUGALL)\renewcommand{\dugall}{DUGALL\xspace}\xspace}
\newcommand{\fwhm}{full-width at half-maximum (FWHM)\renewcommand{\fwhm}{FWHM\xspace}\xspace}
\newcommand{\wht}{William Herschel Telescope (WHT)\renewcommand{\wht}{WHT\xspace}\xspace}
\newcommand{\emccd}{electron multiplying CCD (EMCCD)\renewcommand{\emccd}{EMCCD\xspace}\xspace}
\newcommand{\dasp}{the Durham \ao simulation platform (DASP)\renewcommand{\dasp}{DASP\xspace}\xspace}
\newcommand{\eelt}{European \elt (E-ELT)\renewcommand{\eelt}{E-ELT\xspace}\xspace}
\newcommand{\mpi}{Message Passing Interface (MPI)\renewcommand{\mpi}{MPI\xspace}\xspace}
\newcommand{\smp}{symmetric multi-processing (SMP)\renewcommand{\smp}{SMP\xspace}\xspace}
\newcommand{\svd}{singular value decomposition (SVD)\renewcommand{\svd}{SVD\xspace}\xspace}
\newcommand{\gpu}{graphical processing unit (GPU)\renewcommand{\gpu}{GPU\xspace}\renewcommand{\gpus}{GPUs\xspace}\xspace}
\newcommand{\gpus}{graphical processing units (GPUs)\renewcommand{\gpu}{GPU\xspace}\renewcommand{\gpus}{GPUs\xspace}\xspace}
\newcommand{\fft}{fast Fourier transform (FFT)\renewcommand{\fft}{FFT\xspace}\xspace}
\newcommand{\ifu}{integral field unit (IFU)\renewcommand{\ifu}{IFU\xspace}\xspace}
\newcommand{\darc}{the Durham adaptive optics real-time controller (DARC)\renewcommand{\darc}{DARC\xspace}\renewcommand{\Darc}{DARC\xspace}\xspace}
\newcommand{\Darc}{The Durham adaptive optics real-time controller (DARC)\renewcommand{\darc}{DARC\xspace}\renewcommand{\Darc}{DARC\xspace}\xspace}
\newcommand{\cots}{commercial off-the-shelf (COTS)\renewcommand{\cots}{COTS\xspace}\xspace}
\newcommand{\rtcp}{real-time control pipeline (RTCP)\renewcommand{\rtcp}{RTCP\xspace}\xspace}
\newcommand{\rms}{root-mean-square (RMS)\renewcommand{\rms}{RMS\xspace}\xspace}
\newcommand{\sFPDP}{serial Front Panel Data Port (sFPDP)\renewcommand{\sFPDP}{sFPDP\xspace}\xspace}
\newcommand{\wpu}{wavefront processing unit (WPU)\renewcommand{\wpu}{WPU\xspace}\xspace}

\newcommand{\rtcs}{real-time control system (RTCS)\renewcommand{\rtcs}{RTCS\xspace}\xspace}
\newcommand{\ptp}{point-to-point (PTP)\renewcommand{\ptp}{PTP\xspace}\xspace}
\newcommand{\sse}{streaming SIMD extension (SSE)\renewcommand{\sse}{SSE\xspace}\xspace}
\newcommand{\api}{application programming interface (API)\renewcommand{\api}{API\xspace}\xspace}
\newcommand{\corba}{Common Object Request Broker Architecture (CORBA)\renewcommand{\corba}{CORBA\xspace}\xspace}
\newcommand{\lqg}{linear quadratic gaussian (LQG)\renewcommand{\lqg}{LQG\xspace}\xspace}
\newcommand{\scao}{single conjugate adaptive optics (SCAO)\renewcommand{\scao}{SCAO\xspace}\xspace}
\newcommand{\dma}{direct memory access (DMA)\renewcommand{\dma}{DMA\xspace}\xspace}
\newcommand{\xao}{extreme adaptive optics (XAO)\renewcommand{\xao}{XAO\xspace}\xspace}
\newcommand{\vlt}{Very Large Telescope (VLT)\renewcommand{\vlt}{VLT\xspace}\xspace}
\newcommand{\sparta}{Standard Platform for Advanced Real-Time
  Applications (SPARTA)\renewcommand{\sparta}{SPARTA\xspace}\xspace}
\newcommand{\eso}{European Southern Observatory (ESO)\renewcommand{\eso}{ESO\xspace}\xspace}

\newcommand{\epics}{Exo-Planet Imaging Camera and Spectrograph (EPICS)\renewcommand{\epics}{EPICS\xspace}\xspace}
\newcommand{\iir}{infinite impulse response (IIR)\renewcommand{\iir}{IIR\xspace}\xspace}
\newcommand{\gtc}{Gran Telescopio Canarias (GTC)\renewcommand{\gtc}{GTC\xspace}\xspace}
\newcommand{\cog}{centre of gravity (CoG)\renewcommand{\cog}{CoG\xspace}\xspace}

\title[Lucky imaging with full-sky laser-assisted AO]{Visible near-diffraction
  limited lucky imaging with full-sky laser assisted adaptive optics}

\author[A. G. Basden]{A. G. Basden$^{1}$\thanks{E-mail:
    a.g.basden@durham.ac.uk (AGB)}\\
$^{1}$Department of Physics, South Road, Durham, DH1 3LE, UK}

\ignore{Mag 0 give 2.55 x1e-20 erg/s/cm2/Hz for Iband, 790nm,
  deltalam/lam=0.19.

Or 2550 Janskys

Or 1m phot/s/cm2 in I band.  So 0.064 at 18th Mag.  So, 8143 phot/s on
the WHT pupil.  81 phot/frame at 100Hz.  50percent throughput -> 40phot/frame.
6.4 phot/frame for 20th mag.

}

\begin{document}
\maketitle

\begin{abstract}
Both lucky imaging techniques and adaptive optics require natural
guide stars, limiting sky coverage, even when laser guide stars are
used.  Lucky imaging techniques become less successful on larger
telescopes unless adaptive optics is used, as the fraction of images
obtained with well behaved turbulence across the whole telescope pupil
becomes vanishingly small.  

Here, we introduce a technique combining lucky imaging techniques with
tomographic laser guide star adaptive optics systems on large
telescopes.  This technique does not require any natural guide star
for the adaptive optics, and hence offers full sky-coverage adaptive
optics correction.  In addition, we introduce a new method for lucky
image selection based on residual wavefront phase measurements from
the adaptive optics wavefront sensors.

We perform Monte-Carlo modelling of this technique, and demonstrate
I-band Strehl ratios of up to 35\% in 0.7~arcsecond mean seeing
conditions with 0.5~m deformable mirror pitch and full adaptive optics
sky-coverage.  We show that this technique is suitable for use with
lucky imaging reference stars as faint as Magnitude 18, and fainter if
more advanced image selection and centring techniques are used.
\end{abstract}

\section{The quest for high resolution optical astronomical images:
  Adaptive optics and lucky imaging}
\label{sect:intro}
The lucky imaging concept \citep{1978JOSA...68.1651F}, where selected
short exposure images are integrated into a final image using some
selection criteria, is now an accepted technique in astronomy.  This
concept relies on utilising the rare moments when perturbations
introduced by atmospheric turbulence are minimal, to build up visible
wavelength astronomical images.  The criteria for image selection
(i.e.\ which images to keep for integration, and which to discard) is
a topic of active research
\citep{2012PASP..124..861G,2010SPIE.7735E.196S,2013MNRAS.432..702M}.
However, all rely on having one or more selection stars (guide stars)
within the field of view.  These guide stars are used to determine
whether the selection criterion has been met.  These stars must be
sufficiently bright to ensure that the selection criteria can be
measured within the short atmospheric coherence time, and therefore,
sky coverage is somewhat restricted \citep{2002A&A...387L..21T}.

Lucky imaging works well on 1--2~m class telescopes, where there is
a significant chance of the perturbed wavefront being
instantaneously flat over the whole telescope aperture.  However, on larger
telescopes, the probability of the atmospheric turbulence producing a
flat wavefront over the whole telescope pupil (and hence, a lucky
image) is significantly reduced, with returns reducing as telescope
size increases.  

\Ao \citep{adaptiveoptics} is a technique used to detect time varying
optical perturbations and correct for them in real-time.  Widely used
in astronomy, \ao has led to a breakthrough in ground-based infrared
astronomical imaging, with near diffraction limited images routinely
produced.  However, \ao performance at visible wavelengths is poorer,
with images being far from diffraction limited, though somewhat
improved over uncorrected, seeing-limited images.  A combination of
\ao with lucky imaging has been proposed and implemented
\citep{2012SPIE.8446E..21M,2011MNRAS.413.1524F} with impressive
results.  However, \ao also brings a requirement for sufficiently
bright guide star targets to be used for wavefront sensing, again with
sky-coverage implications.  The use of \lgss \citep{laserguidestar}
with lucky imaging has been proposed \citep{2008SPIE.7015E..67L}, with
the \lgss leading to improved sky-coverage.  Unfortunately, even these
systems require at least one sufficiently bright natural guide star to
overcome the \lgs tip-tilt ambiguity.

Here, we introduce a concept for lucky imaging assisted by \ao where
\ao sky coverage is unlimited.  We propose to use multiple \lgss to
perform tomographic reconstruction of the atmosphere, as is done by
wide-field \ao systems \citep{canaryshort}.  This technique uses no
\ngss and so the reconstructed atmospheric phase contains an unknown
tip-tilt component, translating to an image shift in the telescope
focal plane.  The \ao corrected lucky images are then selected, either
using conventional methods, or using wavefront phase information, a
technique which we introduce and describe here.  Finally, recentring of
the selected lucky images is performed.    

In \S2 a description of these techniques are given, along with details
of modelling and simulations performed.  In \S3 we discuss simulation
results and implications.  In \S4, we discuss plans for on-sky testing
and we conclude in \S5.

\section{Laser guide star assisted lucky imaging}
A tomographic (3D) reconstruction of atmospheric turbulence is
possible when multiple guide stars are available \citep{2007MNRAS.376..287A}.
By correlating wavefront sensor measurements corresponding to
different source directions, the height of the perturbations
introduced by the atmosphere can be determined, and a corresponding
mitigation can be applied.  A requirement for many natural guide stars
has sky-coverage implications.  However, multiple \lgs \ao systems
have recently come online \citep{2012SPIE.8447E..0KM}, allowing tomographic
reconstruction to be performed with only one natural guide star (which
is required to determine the global tip-tilt, to which \lgss are insensitive).

\subsection{Using wavefront phase for image selection}
A projection of the reconstructed turbulent volume along any given
line of sight allows the perturbed wavefront incident from any
direction within the field of view to be estimated, as has been
successfully demonstrated by \moao systems
\citep{2007MNRAS.376..287A,basden12}.  This information can then be
used as a criterion for lucky image selection: If the \rms wavefront
error is below a particular value at a given instant in time then the
corresponding lucky image should be used rather than discarded.  This
has an advantage over traditional means of image selection in that
since the perturbed wavefront can be estimated over the entire field
of view, it is now possible to select only parts of this field for
integration, while discarding parts where wavefront perturbation is
higher.

\subsubsection{Laser guide star tip-tilt measurement}
If \lgs information is used to compute the selection criteria metric
for lucky imaging as previously described, the unknown tip-tilt
component (due to the lack of \ngs information) manifests itself as
common lateral movement of objects in the image plane.  Therefore,
individually selected lucky images can simply be shifted and added.
However, unless taken into account, this unknown wavefront tilt will
affect the selection criteria which is based on wavefront error, since
rms wavefront error increases with wavefront tilt.  Therefore, it is
necessary to remove the tip and tilt components from the reconstructed
wavefronts before using them to determine whether the image selection
criteria have been met.  In other words, {\em using \lgs tomography, a
  selection criteria for lucky images can be obtained, based on the
  rms of tilt-removed reconstructed wavefront phase in any given
  source direction}.

These selected images can then be shifted and added, with the shift
determined either from a natural guide star, or from information
within the lucky image itself, for example, the centroid location of a
bright star or Fourier information across the whole image (since the
tip-tilt can be assumed to be global over a few tens of arcseconds).
In the case where selection criteria is based on wavefront error from
the \lgss, it should be noted that since image selection is not
performed using the individual lucky images, reduced photon flux
within these images may be acceptable, as we will show in following
sections.  It should also be noted that centroiding of any lucky
images requires enough light within the images in the field
to be present, and that this is generally not effective on diffuse
objects such as galaxy images.

\subsection{Wavefront correction with a deformable mirror}
\ao entails the correction of a wavefront using a \dm.  This can be
combined with lucky imaging for improved performance
\citep{2012SPIE.8446E..21M}.  The combination of lucky imaging with a single
\lgs \ao system has also been suggested and demonstrated
\citep{2008SPIE.7015E..67L}.  However, in this case, the lucky image
selection criteria was determined from the lucky images, and a \ngs
was required to avoid tip-tilt ambiguity for the \ao correction, and
thus sky coverage was limited.

This \ngs requirement is however not necessary if the \ao correction
is aiming to only sharpen each individual lucky image rather than the
mean long exposure image.  In this case, using multiple \lgss such
that the turbulent volume is well sampled, an \ao correction can be
computed with the tip/tilt component present, but ignored.  The
corrected \ao image will then have an undetermined instantaneous tip
and tilt (position shift in the image plane), but be otherwise
relatively well corrected.  The selected lucky images (using either
the wavefront phase criterion estimated from \lgs information, or
conventional image selection techniques) can then be shifted and
added, with the shift determined using the information within the
lucky image (i.e.\ light from all sources within limits defined by the
field of view or the tip-tilt isoplanatic patch size).  It should be
noted that markedly poorer resolution can be obtained if image
recentering on multiple targets is not performed with care,
essentially due to tip-tilt anisoplanatism, causing radial scale
variations.

If the \lgss are operating in open-loop, i.e.\ are insensitive to
changes to the \dm surface (as in the case of a \moao system such as
CANARY), then there is a degree of flexibility in the mode of \ao
correction that is performed, including \moao (if several \dms are
available), \ltao (to correct a single line of sight) and \glao (to
perform moderate correction over a wide field of view).

\subsubsection{Estimation of open-loop AO corrected image phase}
If the \lgss are operating in open-loop, then the estimated wavefront
phase will not be equal to that seen by the lucky imaging detector,
since this is placed after the \dm.  However, as we will show, it is
still possible to use the reconstructed phase to derive the lucky
image selection criterion.

\subsection{Modelling of LGS assisted lucky imaging}
We have used \dasp \citep{basden5,basden8} to perform full end-to-end
Monte Carlo simulation of \lgs assisted lucky imaging, and here
describe our model.  This is based loosely around the CANARY
instrument, since we intend to use this instrument for on-sky testing
of the concepts described in this paper.  We assume a 4.2~m telescope
with a 1.2~m central obscuration (representative of the William
Herschel Telescope).  We use four Rayleigh \lgss at 22~km, placed in a
circular asterism with a 40~arcsecond diameter, and cone effect, spot
elongation and laser uplink effects are included in our simulations.
A 532~nm wavelength is assumed for the \lgs.  For comparison, we also
show results when sodium \lgss are used, with the sodium beacon at
90~km.  

When performing \ao correction, the \dm is operated in open-loop
(i.e.\ not sensed by the \wfss).  We vary the number of sub-apertures
between $8\times8$ (the default case) and $32\times32$, with 4 pixels
per sub-aperture, and a pixel scale of 0.84~arcseconds per pixel.  We
model a \dm with $5\times5$ and $9\times9$ actuators, with $9\times9$
being the default case (which is to be assumed unless stated
otherwise).  We use a 9 layer atmospheric profile \citep{basden12}
with a 20~m outer scale and a Fried's parameter, $r_0=13.5$~cm,
corresponding to 0.8~arcsecond seeing, unless otherwise stated.  The
simulation time-step is 2.5~ms, the \wfs exposure time is 5~ms (2
simulation time-steps), and the lucky image exposure time is 10~ms (4
simulation time-steps).  We simulate 250~s of telescope time for each
case that is considered.

We ignore the mean tip-tilt component of each \lgs when reconstructing
wavefront phase perturbations, and do not use any additional tip-tilt
information, i.e.\ tip-tilt remains uncorrected.  Wavefront
reconstruction is performed by placing virtual \dms at each turbulent
layer, and using a minimum variance reconstruction formulation
\citep{map} to derive the corresponding wavefront phase.  We then
propagate the tomographically reconstructed wavefront phase along a
given line of sight to estimate the wavefront perturbation in that
particular direction.  

\begin{figure}
(a)\vspace{0.4cm}\\
\includegraphics[width=0.8\linewidth]{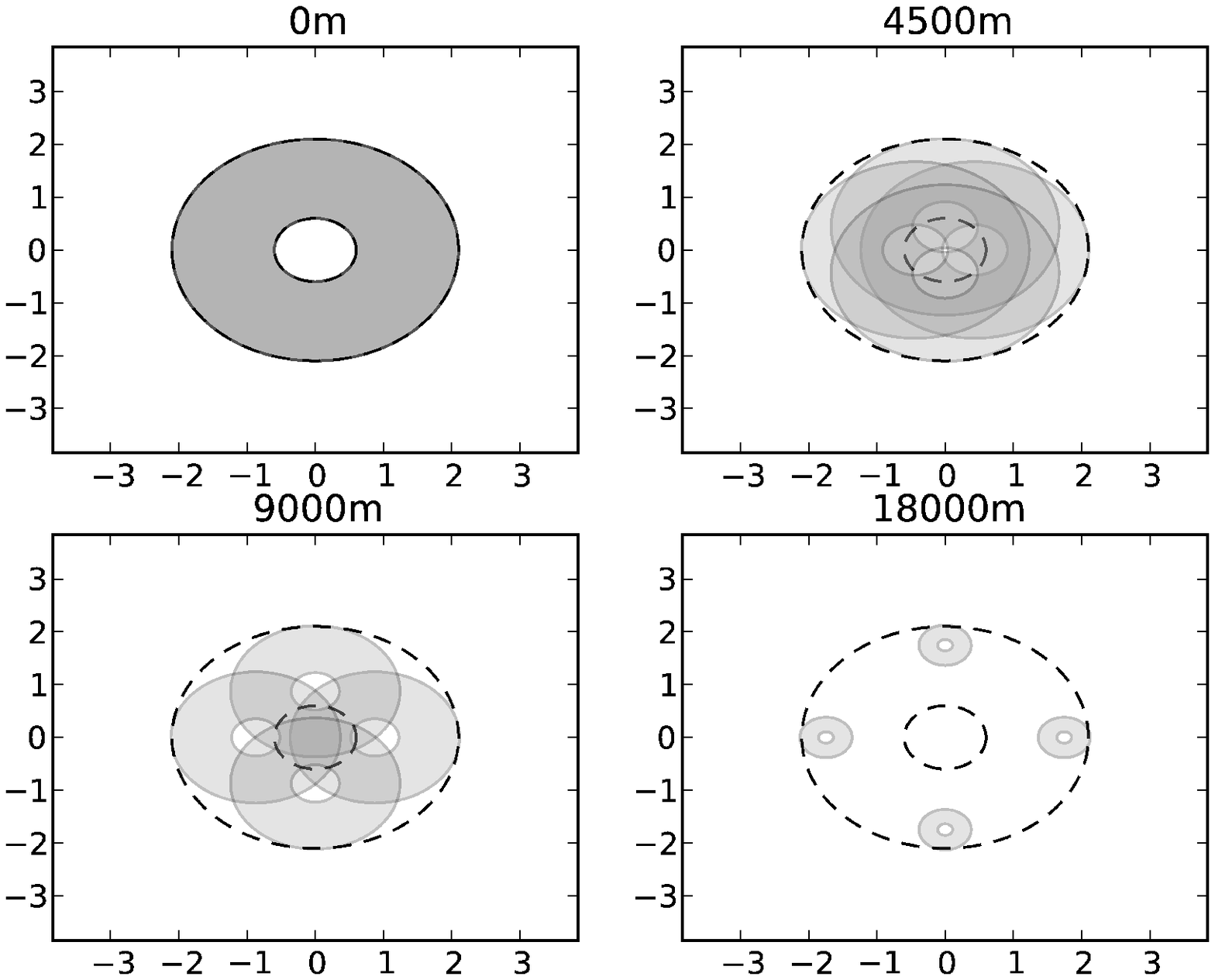}\vspace{0.2cm}\\
(b)\vspace{0.4cm}\\
\includegraphics[width=0.8\linewidth]{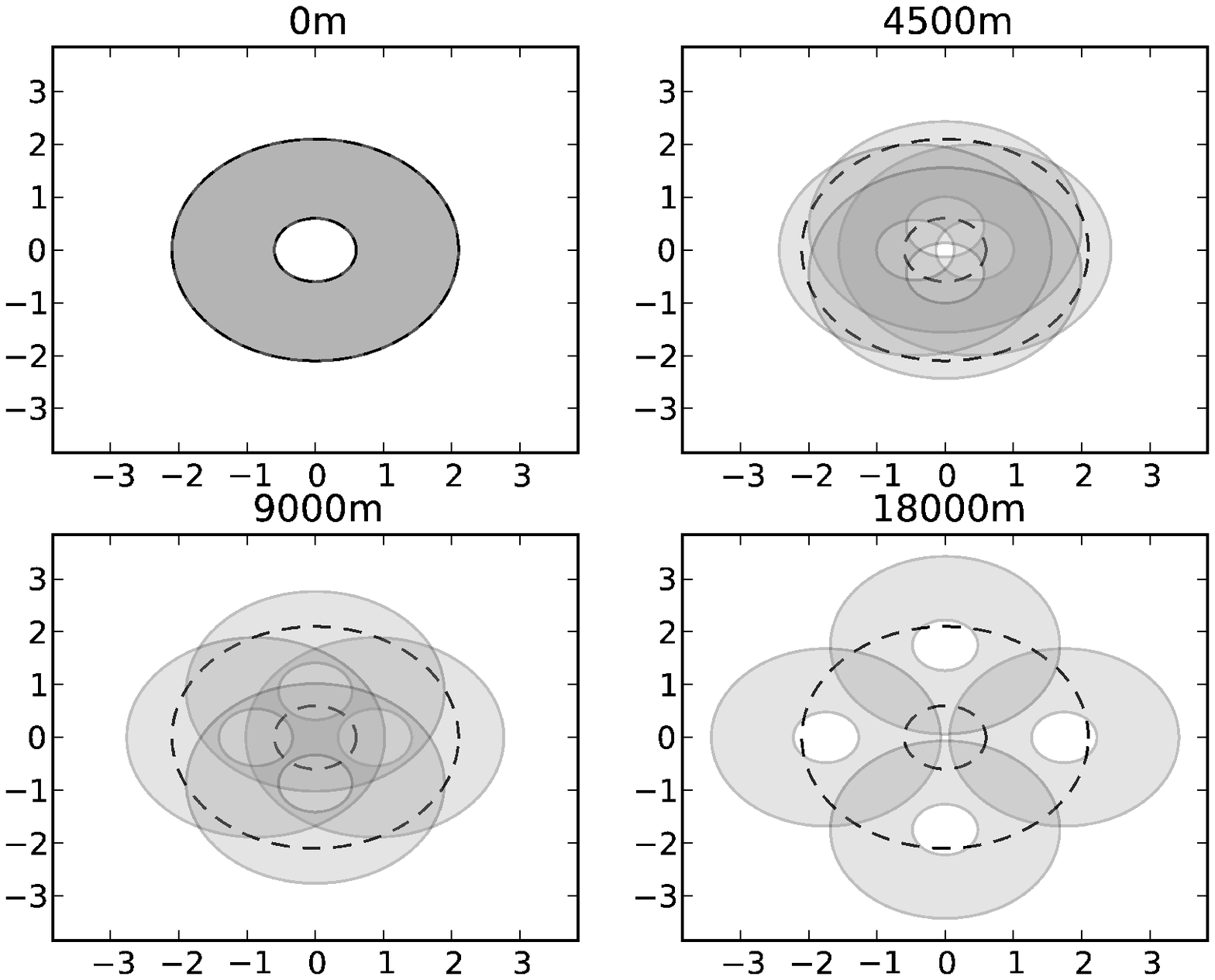}
\caption{A figure showing LGS pupil overlap at different heights
  (shown in the titles) above the telescope.  (a) For Rayleigh LGS
  focused at 22~km.  (b) For sodium LGS focused at 90~km.}
\label{fig:asterism}
\end{figure}

Fig.~\ref{fig:asterism} shows the \lgs pupil overlap at the heights of
four atmospheric layers: Tomographic reconstruction of wavefront phase
is theoretically possible whenever two guide star pupils overlap, and
therefore the 90~km case provides better coverage of tomographic phase
information.

The lucky images are generated assuming a detector with $256\times256$
pixels and a field of view of 5~arcseconds for each line of sight.
The angular resolution of the telescope at the 800~nm wavelength used is about
0.05~arcseconds.  We consider lucky images for sources up to one
arcminute off-axis, with on-axis results being presented by default.
These images are considered to have high signal-to-noise ratio (no
noise) unless otherwise stated.

We assume idealised telescope tracking and stability, and that there
the non-common path errors between the \ao system and the lucky
imaging system have been removed.  We also assume that the telescope
is well focused.  Real telescope operational parameters will result in
slight performance degradation, particularly when operating without
the \ao loop.  This will be of particular importance when there are
high frequency vibration modes within the telescope structure, at
frequencies approaching that of the lucky imaging system frame rate.
However, on the \wht, we have identified the highest frequency
vibration with significant power at 22~Hz \citep{lqgshort}, which will
have negligible impact on individual lucky image quality, and can be
corrected using vibration control techniques, such as a Kalman filter,
as we have demonstrated \citep{lqgshort}.  However, the effect of these
assumptions should be borne in mind.

\subsubsection{Variation of $r_0$}
When used on-sky, lucky imaging takes advantage of the fact that the
instantaneous atmospheric Fried's parameter, $r_0$, is not constant.
Therefore at times when $r_0$ is large, good images are collected,
while when $r_0$ is smaller, more images are thrown away.  In our
simulations, we usually assume that $r_0$ is constant, meaning that
our results will be somewhat pessimistic.  We do, however, provide
results from simulations performed with different values of $r_0$.
Additionally, we also include results from simulations where the value
of $r_0$ is changing throughout the simulation period.  

\subsection{Lucky selection criteria}
For comparative purposes, we use selection criteria derived from the
lucky images themselves, and from the tomographically projected (along
each line of sight of interested) tilt removed rms wavefront
perturbation (with flatter wavefronts producing cleaner \psfs).  

Selection criteria derived from the lucky images themselves include
Strehl ratio and diameter encompassing 50\% of image energy.  It
should be noted that using a Strehl ratio selection criteria
introduces a performance bias when Strehl ratio is also used to
determine final image quality, leading to overestimation of
performance, particularly at low signal levels when noise can have a
large effect on measured instantaneous Strehl ratio.  However, we find
it useful to include these results nevertheless, as image selection by
Strehl ratio is conceptually easy to understand.

\subsection{Shift and adding images}
Since the \lgs do not measure, and cannot correct, mean image
position, we have to recenter the selected lucky images before
summation.  There are many ways to do this in the literature
\citep{2013MNRAS.432..702M}.  Here, we simply measure the image centre
of gravity, and shift the image by a whole number of pixels.  Although
this is not optimal, it maintains simplicity in our results.  

The final images (comprising many shifted and added selected lucky
images) are then processed to compute Strehl ratio, \fwhm, and
diameter encircling 50\% of energy.  We do not consider any Fourier
based image selection techniques here.

\section{Performance of LGS assisted lucky imaging}
\begin{figure}
\includegraphics[width=\linewidth]{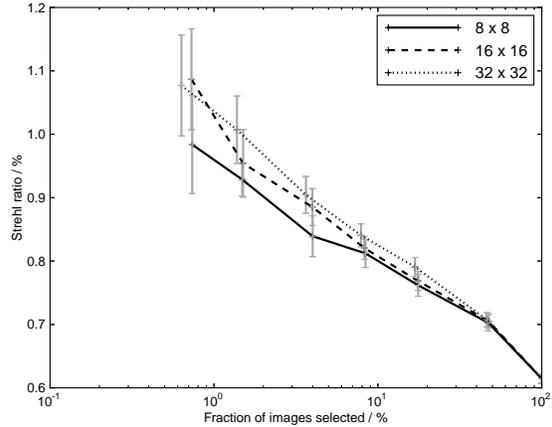}
\caption{A figure showing lucky image quality (Strehl ratio) as a function of
  fraction of images selected.  The solid curve represents selection
  using $8\times8$ sub-aperture WFSs, the dashed curve is for
  $16\times16$ sub-apertures and the dotted curve is for $32\times32$
  sub-apertures.  No AO correction is performed.}
\label{fig:uncorr}
\end{figure}

We consider first the case where no \ao correction is performed, to
investigate the \lgs wavefront phase based selection criteria.  Here,
we use the \lgss to tomographically compute wavefront phase along a
given line of sight, and the lucky image selection criteria is derived
from the rms wavefront phase.  Fig.~\ref{fig:uncorr} shows on-axis
lucky image quality as a function of fraction of images selected, for
different \wfs orders (number of sub-apertures), which are used to
obtain the lucky image selection criteria.  It should be noted that no
\ao correction is performed, i.e.\ the \lgss are used for wavefront
measurement only.  As expected, the image quality is reduced as a
greater fraction of images are selected.  It can also be seen that
using a higher order \wfs generally gives a better image selection
measurement, as would be expected due to the increased detail in
reconstructed phase.  The uncorrected, long exposure Strehl ratio in
this case is about 0.5\%, and therefore a factor of two improvement
can be achieved.

\subsection{Adaptive optics correction using LGS measurements}
\begin{figure}
\includegraphics[width=\linewidth]{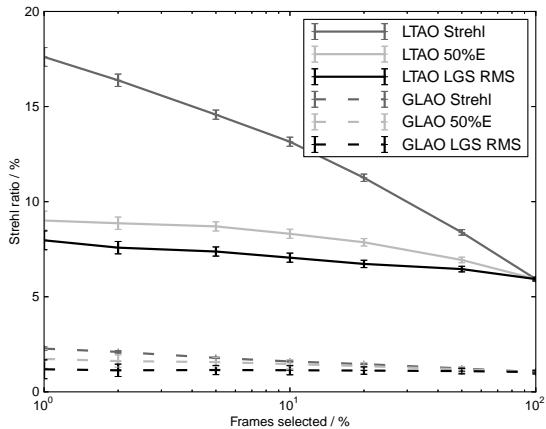}
\caption{A figure showing integrated lucky image quality as a function
  of fraction of frames selected, for selection criteria based on \lgs
  rms wavefront error, on image Strehl ratio and on PSF diameter
  encircling 50\% of energy (50\%E).  Cases using both
  a \ltao correction and a \glao correction are shown (both
  without global tip/tilt correction.}
\label{fig:glao}
\end{figure}

We now consider the case where \ao correction is performed, using a
\dm, the surface of which is shaped using the \lgs \wfs measurements.
As detailed in the previous section, the mean \lgs tip-tilt signal
from each \lgs is ignored, such that the \ao correction improves the
instantaneous \psf, but this is not in a fixed location.  Therefore,
shifting and adding of selected lucky images is still required.  We
consider the case where a tomographic \glao correction is performed
(i.e.\ the \dm is used only to correct ground layer turbulence), and
also where a \ltao correction is performed (i.e.\ the tomographic
wavefront estimate is projected along the on-axis line of sight).
Fig.~\ref{fig:glao} shows that \ltao correction offers significant
performance benefits for on-axis lucky imaging over a global \glao
correction.  This also shows that lucky image selection using \lgs
wavefront rms leads to poorer integrated images than when Strehl ratio
is used as the selection criteria, though does lead to some image
quality improvement.  Using \psf diameter as the lucky selection
criteria gives performance close to that obtained using rms \lgs
wavefront error.

\subsubsection{Open-loop WFS selection criteria}
In the simulations presented here, our \wfss are operating in
open-loop, so that they are insensitive to \dm changes.  Therefore the
selection criteria based on reconstructed phase requires adjustment so
that the \dm correction is taken into account.  We achieve this by
computing the pseudo-closed-loop \wfs signals, and using these to
compute residual (corrected) phase, and hence selection criteria.

\begin{figure}
\includegraphics[width=\linewidth]{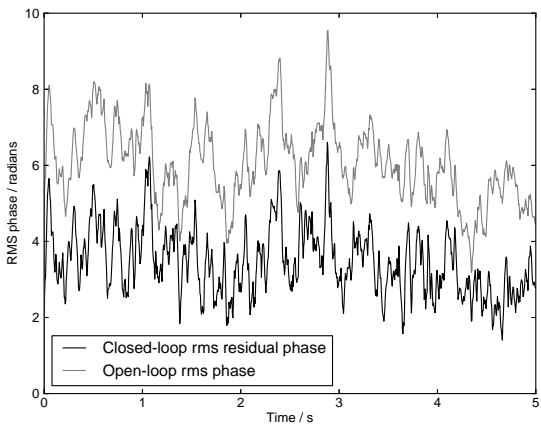}
\caption{A figure showing the rms reconstructed tip-tilt removed
  wavefront phase as a function of time using open-loop WFSs, and
  pseudo-closed-loop WFSs.}
\label{fig:residuals}
\end{figure}

As shown in Fig.~\ref{fig:residuals}, there is a high degree of
correlation between the computed residual wavefront
phase rms and the open-loop uncorrected wavefront phase
rms, though with a residual rms having a consistently lower value (due
to removal of lower order turbulent modes by the \dm).  It is
therefore sufficient to use the open-loop reconstructed
wavefront phase rms as the lucky image selection criteria since a
constant offset in criteria does not affect the images selected (we
select a constant fraction).  

\ignore{
A comparison of lucky image quality when
using open-loop and pseudo-closed loop rms wavefront phase for image
selection is shown in Fig.~\ref{fig:residualComparison}.  For the
remainder of this paper, we use the open-loop variant, to maintain
consistency with lucky images produced without \ao correction, and to
simplify on-sky operation.
}

\ignore{
\begin{figure}
TODO: Run getClosedloopSummary.py on gpu2.
\caption{A figure showing lucky image quality as a function of
  fraction of images selected, comparing selection derived from the
  open-loop and pseudo-close-loop WFS measurements.  For this figure,
  the WFSs had $8\times8$ sub-apertures, and a $9\times9$ actuator
  DM.}
\label{fig:residualComparison}
\end{figure}
}

\subsubsection{LGS assisted lucky imaging performance}
\begin{figure}
\includegraphics[width=\linewidth]{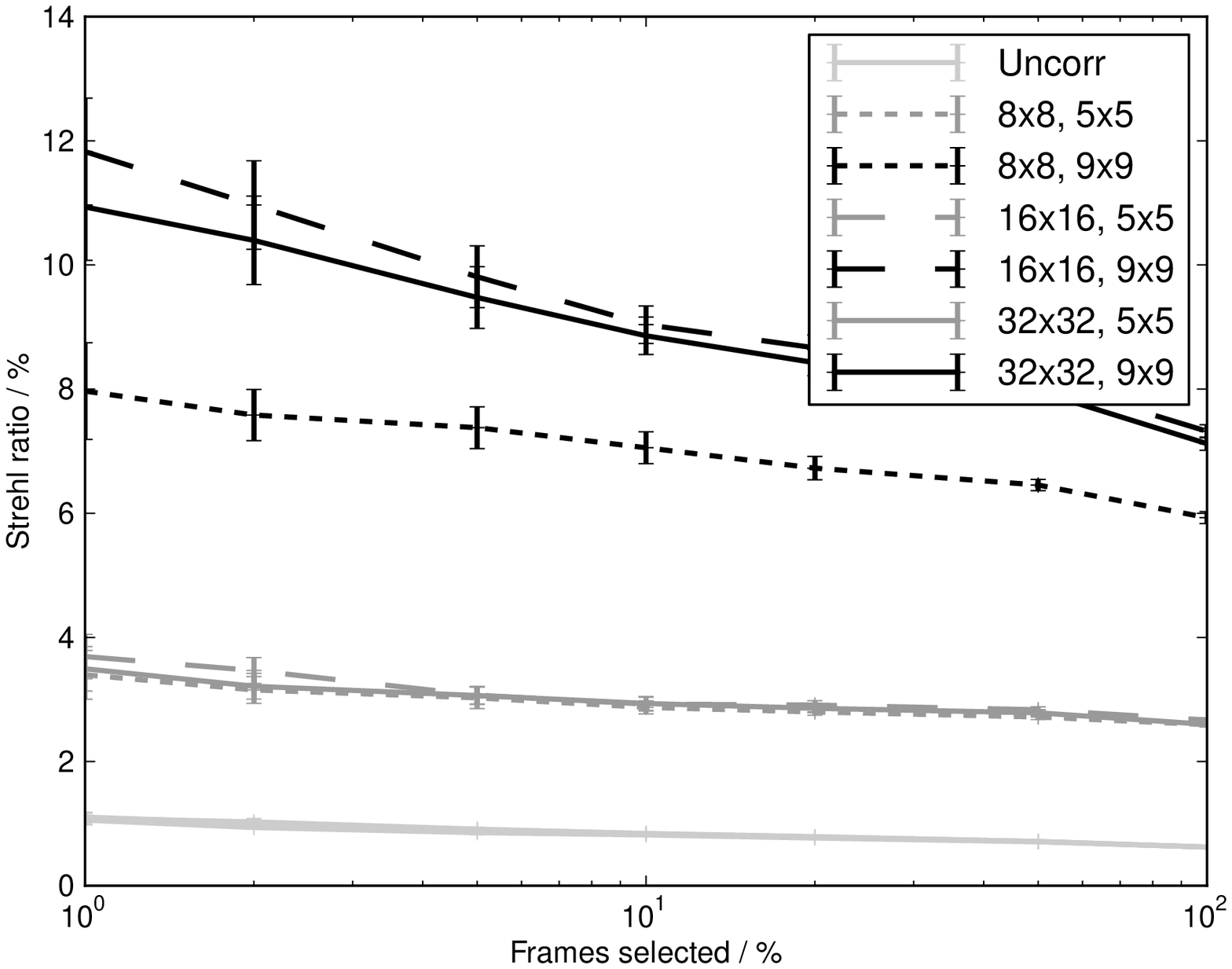}
\includegraphics[width=\linewidth]{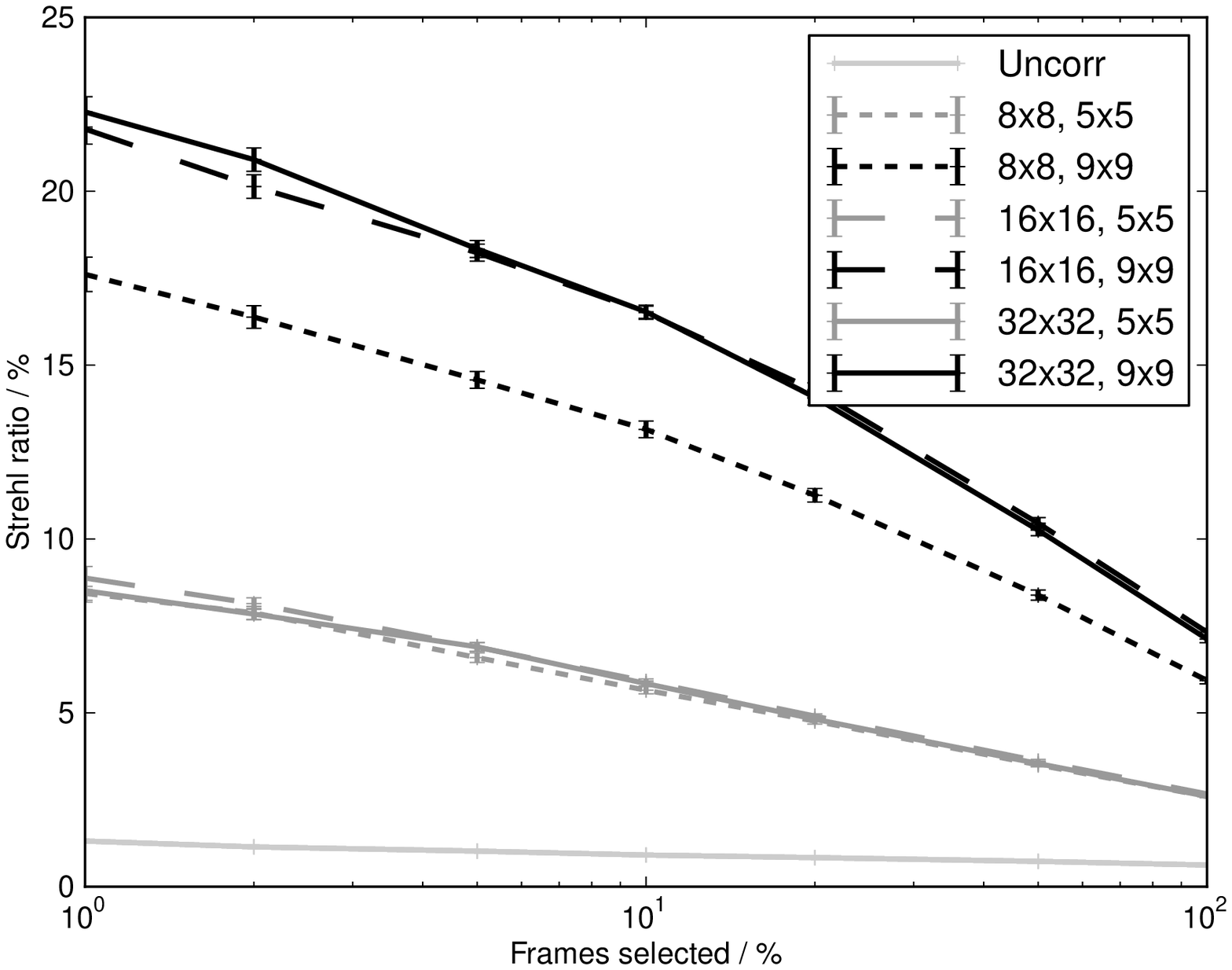}
\caption{(a) A figure showing lucky image quality as a function of
  fraction of images selected, comparing different WFS and DM order as
  given in the legend (an even number of WFS sub-apertures, and an odd
  number of DM actuators).  Uncorrected (no AO) performance is shown
  for comparison, and is relatively independent of WFS order (three
  cases are shown, but overlap, and hence only a single legend entry
  is given).  (b) As for (a), but with the lucky image selection
  criteria derived from the short exposure Strehl ratios.}
\label{fig:ltao}
\end{figure}

Fig.~\ref{fig:ltao}(a) shows lucky image quality as a function of
fraction of images selected, for selection criteria based on
reconstructed wavefront.  We consider several cases for \wfs and \dm
order, and the uncorrected case (without \ao) is shown for comparison.
It should be noted that a higher \wfs order does not necessarily lead
to better image selection, as using $16\times16$ sub-apertures gives a
similar final image Strehl ratio to $32\times32$ sub-apertures (since a
\dm with fewer actuators is used), i.e.\ it is not necessary to
significantly over-sample the \wfs with respect to the \dm, though
some over-sampling does improve \ao correction (when flux is not
limited), as shown by the $16\times16$ sub-aperture case giving better
performance than the $8\times8$ sub-aperture case.  

Fig.~\ref{fig:ltao}(b) shows improved lucky image quality when lucky
image Strehl ratio is used as the image selection criteria, and when
the \wfs order is increased above the \dm order.  This represents the
extra information available during wavefront reconstruction being used
to perform better \ao correction.

\subsubsection{Lucky image noise considerations}
We now consider the case where noise is present within the lucky
images, investigating a range of signal levels and readout noise.
Lucky images are usually captured on flux-limited targets (due to the
necessity for high frame rate), using low light level detectors, with
photon counting strategies \citep{basden1}.  This has two effects.  In
the case where the image selection criteria is derived from the lucky
image, the noise will affect the accuracy of image selection.
Additionally, the noise will affect the accuracy of mean tip-tilt
calculation (which is taken from the image) when shifting and adding
the selected images.

We introduce photon shot noise, and also detector readout noise of
both 0.1 electrons (corresponding to an \emccd) and one electrons
(corresponding to a sCMOS detector), for the lucky imaging detectors.
In the results that we present, it should be noted that we do not take
detector quantum efficiency into account, nor the excess noise factor
that affects \emccd technologies (effectively halving quantum
efficiency).  Rather, we present results with detected signal level,
giving the reader the freedom to scale these to incident flux
depending on situation.

\begin{figure}
\includegraphics[width=\linewidth]{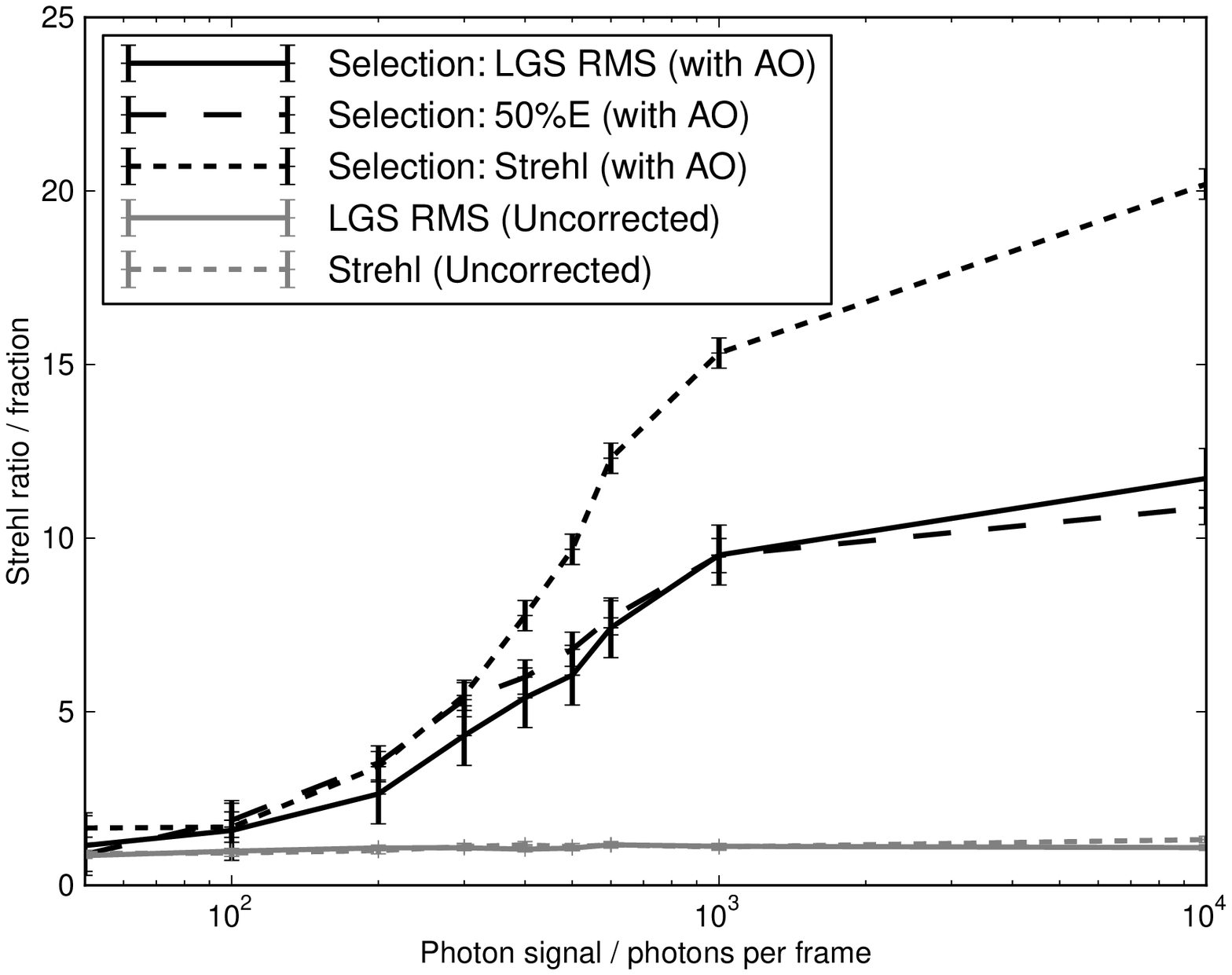}
\caption{A figure showing lucky image quality as a function of
  detected signal
  level for selection criteria derived from rms wavefront and from
  individual image Strehl ratio, for a detector with 0.1 electrons
  readout noise.}
\label{fig:noise}
\end{figure}

Fig.~\ref{fig:noise} shows lucky image quality as a function of signal
level (in detected photons per lucky image frame), for image selection based both on the
\lgs signal and on lucky image characteristics (Strehl ratio and \psf
diameter encircling 50\% energy).  It can be seen that using image
Strehl as the selection criteria gives better performance at all light
levels.

\begin{figure}
\includegraphics[width=\linewidth]{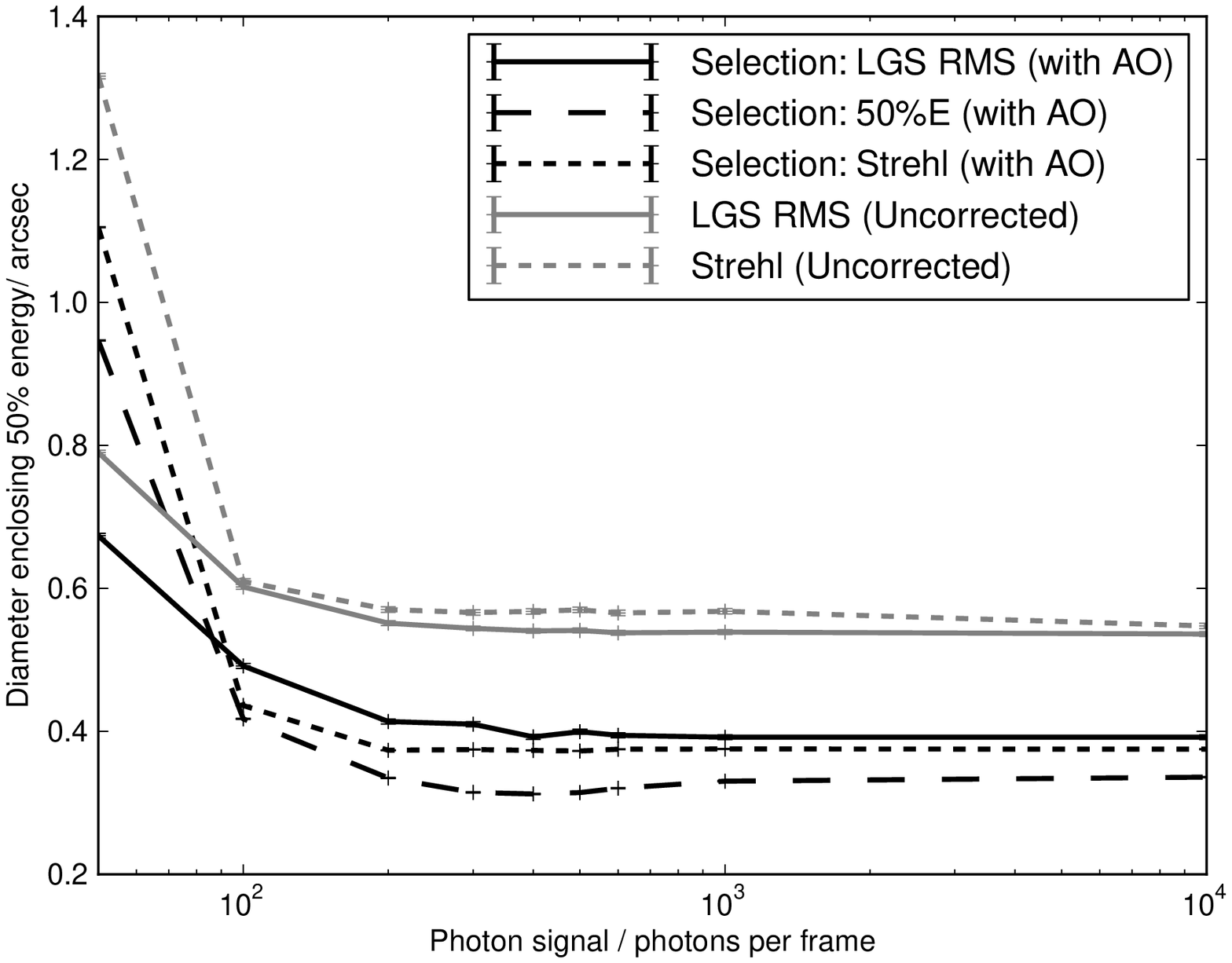}
\includegraphics[width=\linewidth]{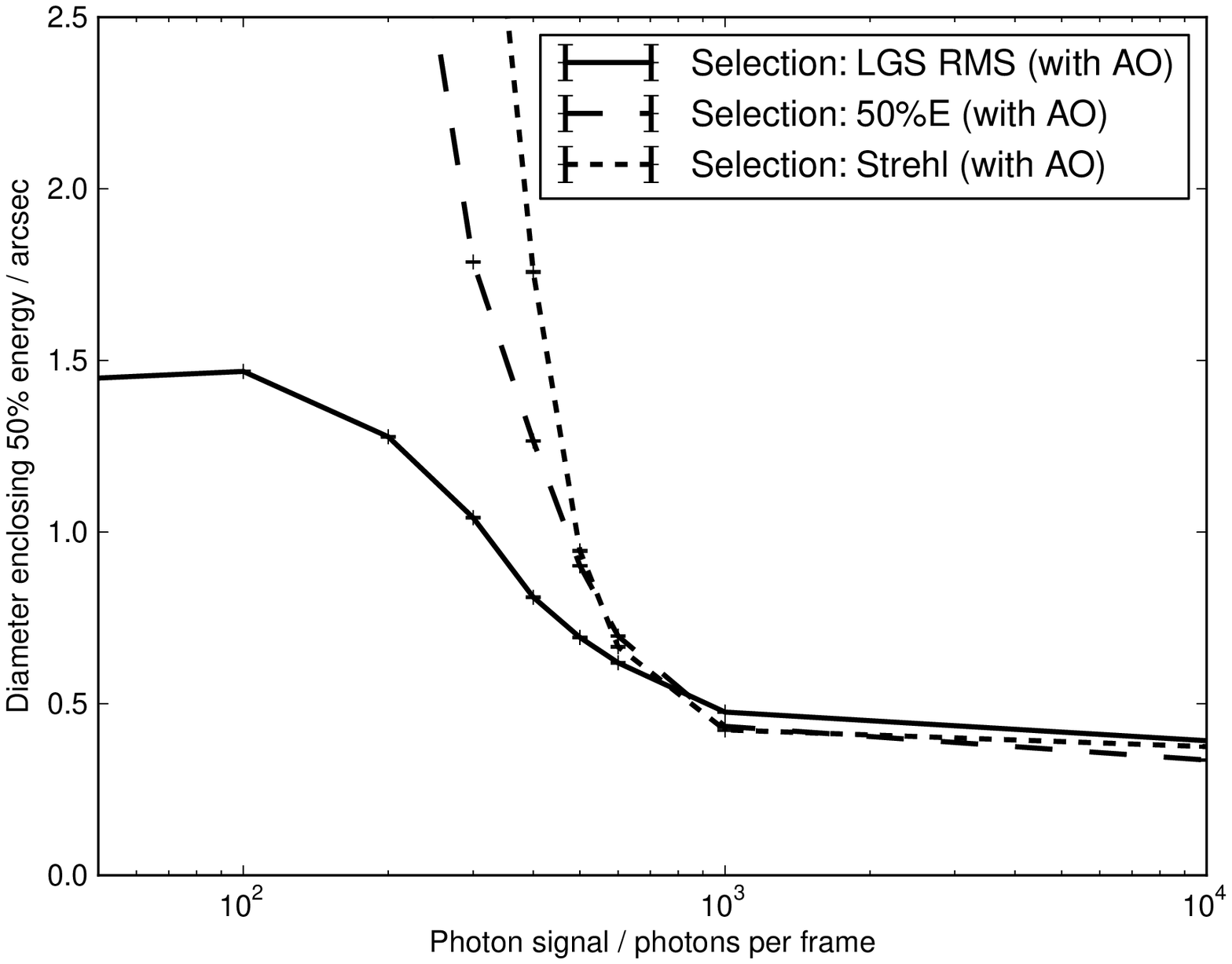}
\caption{A figure showing integrated lucky PSF image diameter
  encircling 50\% of energy as a function of detected signal level for
  selection criteria derived from rms wavefront and from individual
  image Strehl ratio.  A smaller diameter represents better
  performance.  (a) For a detector with 0.1 electrons readout noise.
  (b) For a detector with 1 electron readout noise.}
\label{fig:noised50}
\end{figure}

Many science cases require images defined by criteria other than
Strehl ratio, including for example ensquared energy within some box
size (usually for a spectrograph), or the \psf diameter encircling
some fraction of the available energy.  Fig.~\ref{fig:noised50} shows
lucky image quality based on \psf diameter encircling 50\% of available
energy, for different detector readout noise, using rms wavefront
(from the \lgs), lucky image Strehl ratio and \psf diameter as
selection criteria.  It can be seen here that at lower light levels,
using the \lgs rms wavefront as a selection criteria gives better
performance than selection using the lucky images.  The reason for
this is that at these light levels, the estimated Strehl ratio will be
affected significantly by photon shot noise and readout noise, which
will significantly increase the probability of erroneous image
selection.  When considering Strehl ratio as the final image quality
metric, this erroneous selection will have less of an effect than when
other metrics, such as encircled energy diameter, are used.  For the
case of Strehl ratio, a bright pixel (amplified for example by
unlikely Poisson or Gaussian statistics) will be identified as
corresponding to a high Strehl ratio, and thus be selected.  This
lucky image will then be re-centred and the bright pixel will improve
Strehl of the final image.  However, this instantaneous image could
well have a broad \psf, with a wide energy spread, which will increase
\psf diameter, resulting in poorer image quality, even though Strehl
ratio is (spuriously) higher.  It is therefore necessary, in the case
of low photon flux, to consider carefully the metric used to measure
image quality.

From Fig.~\ref{fig:noise}, it can be seen that there is little
performance improvement once signal levels reach about 1000 detected
photons per lucky image frame.
This corresponds to a I-mag 15 star on a 4.2~m telescope with a 100~Hz
frame rate \citep{1979PASP...91..589B}.  Likewise,
Fig.~\ref{fig:noised50}(a) shows that only 100 photons per lucky image
(with a readout noise of 0.1 electrons) are required before little
additional performance improvements is gained in the case when the
integrated image quality metric is \psf diameter, corresponding to an
I-mag 18 star.  The presence of additional stars within the
field-of-view would allow further improvements to be made, as would
more advanced image selection and centring approaches, though we do
not discuss this further here, rather relying on a simple
centre-of-gravity calculation for image recentring.

\subsection{Performance with different $r_0$}
As previously discussed, lucky imaging relies on rapid variability of
$r_0$ to work well, using images selected when $r_0$ is larger.  We
now consider four different situations in our simulations:  Constant
values of $r_0$ of 13.5~cm, 15~cm and 20~cm, and 
when $r_0$ varies sinusoidally from 10-20~cm with a 12.5~s period
(corresponding to a mean $r_0$ of 15~cm for a 250~s simulation).

\begin{figure}
\includegraphics[width=\linewidth]{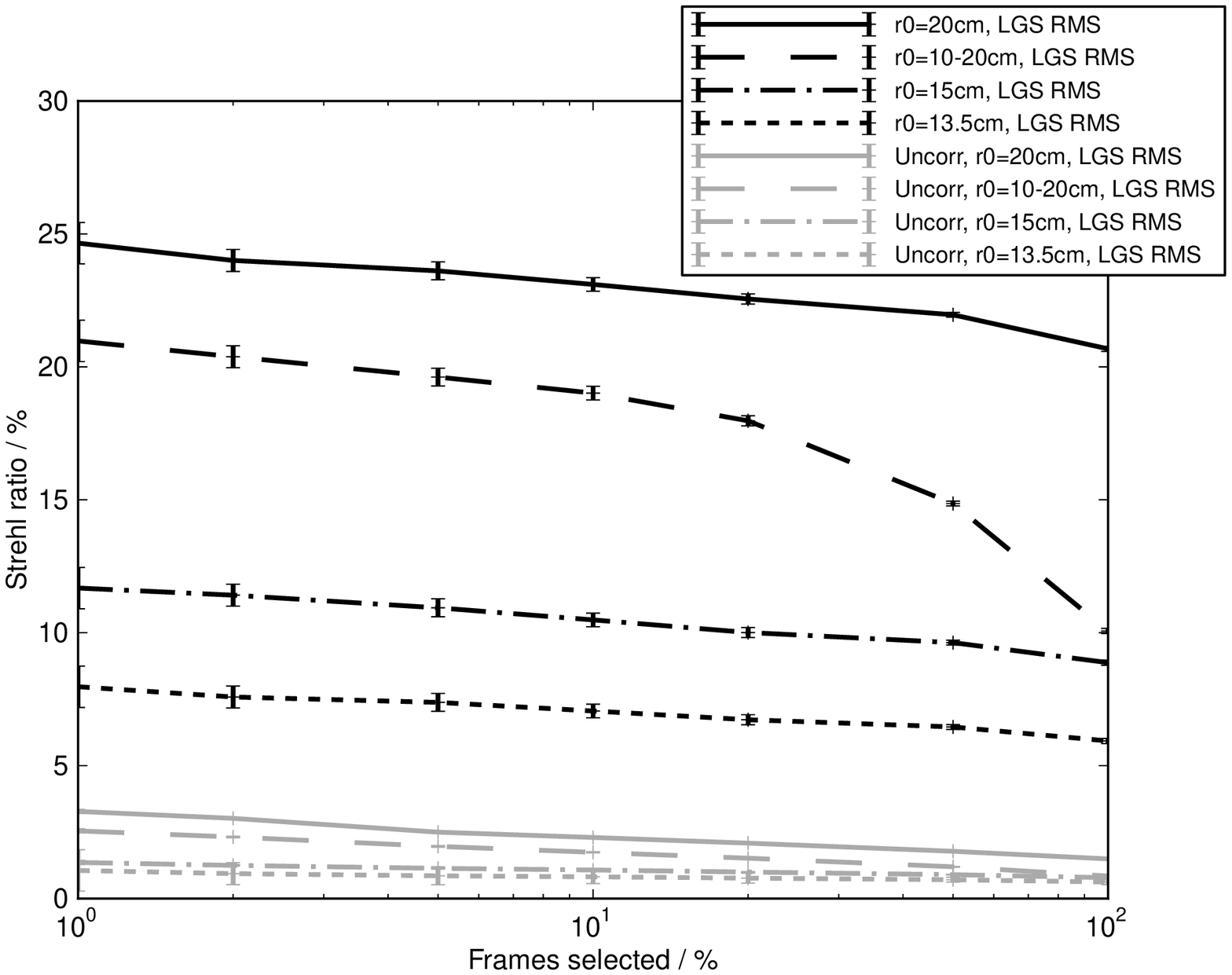}
\caption{A figure showing lucky image quality as a function of
  fraction of images selected for different atmospheric Fried's
  parameters as given in the legend, with rms wavefront phase used for
  image selection.  The AO corrected and non-corrected cases are both shown
(with ``Uncorr'' in the legend representing the un-corrected cases).}
\label{fig:r0}
\end{figure}

Fig.~\ref{fig:r0} shows lucky image quality as a function of image
selection fraction for these different cases, when the selection
criteria is derived from rms \lgs wavefront error.  As expected, a
larger value of $r_0$ results in better performance.  It can also be
seen that in the case of a varying $r_0$, good performance is obtained
when selection fraction is low (since the instantaneous images
selected are generally those obtained during periods of high $r_0$),
while image quality falls rapidly when the fraction of selected images
is increased, since short exposure images captured during worse seeing
conditions then have to be used.  This shows that simulations of lucky
imaging which use a constant $r_0$ value will be pessimistic.
Corresponding uncorrected, long exposure Strehl ratios are 1.2\%
(20~cm $r_0$), 0.71\% (Variable $r_0$), 0.62\% (15~cm $r_0$) and 0.5\%
(13.5~cm $r_0$).

\begin{figure}
\includegraphics[width=\linewidth]{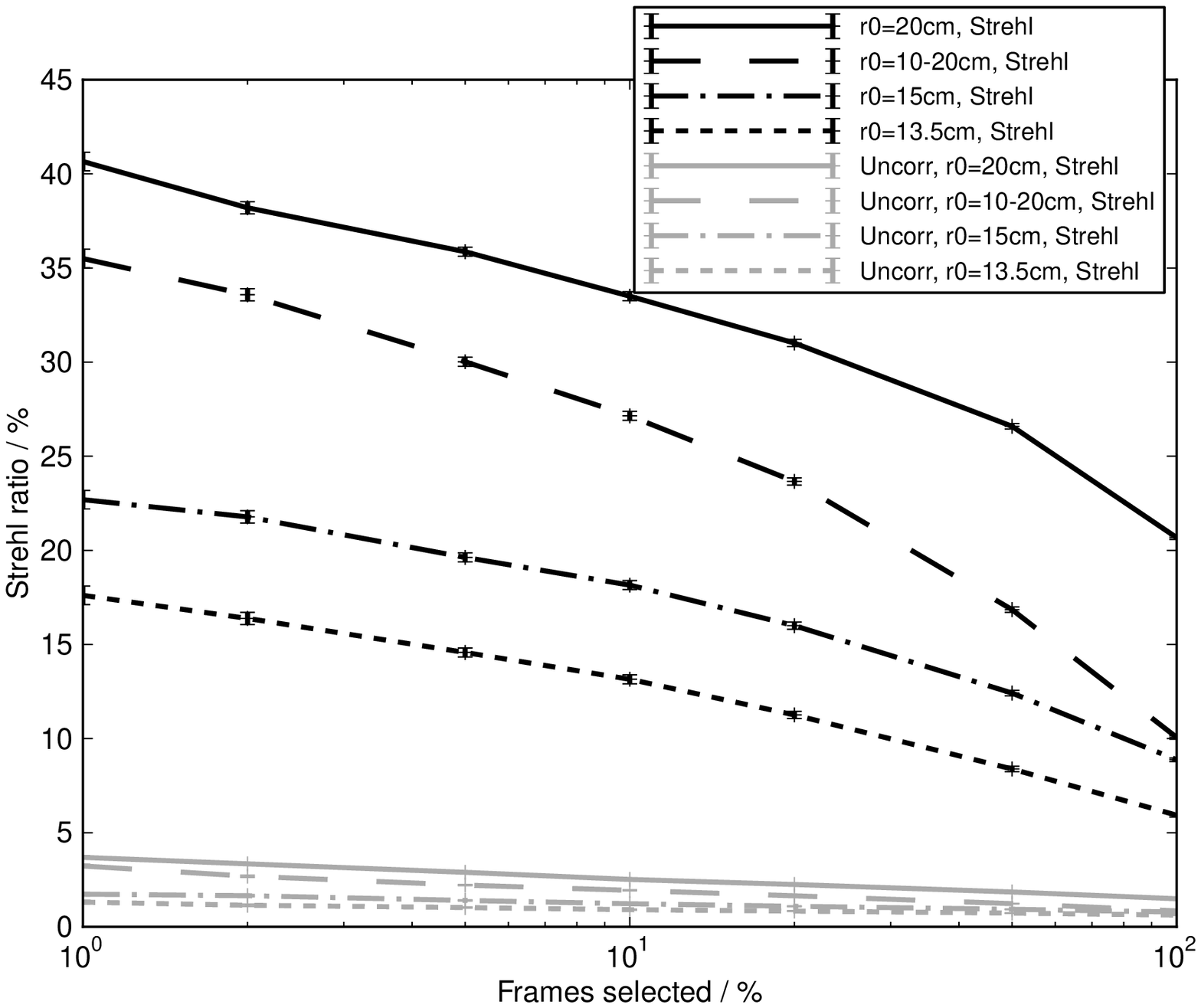}
\caption{A figure showing lucky image quality as a function of
  fraction of images selected for different atmospheric Fried's
  parameters as given in the legend, with Strehl ratio used for image
  selection.  The AO corrected and non-corrected are shown, with
  ``Uncorr'' representing the un-corrected cases.}
\label{fig:r0strehl}
\end{figure}

\begin{figure}
\includegraphics[width=\linewidth]{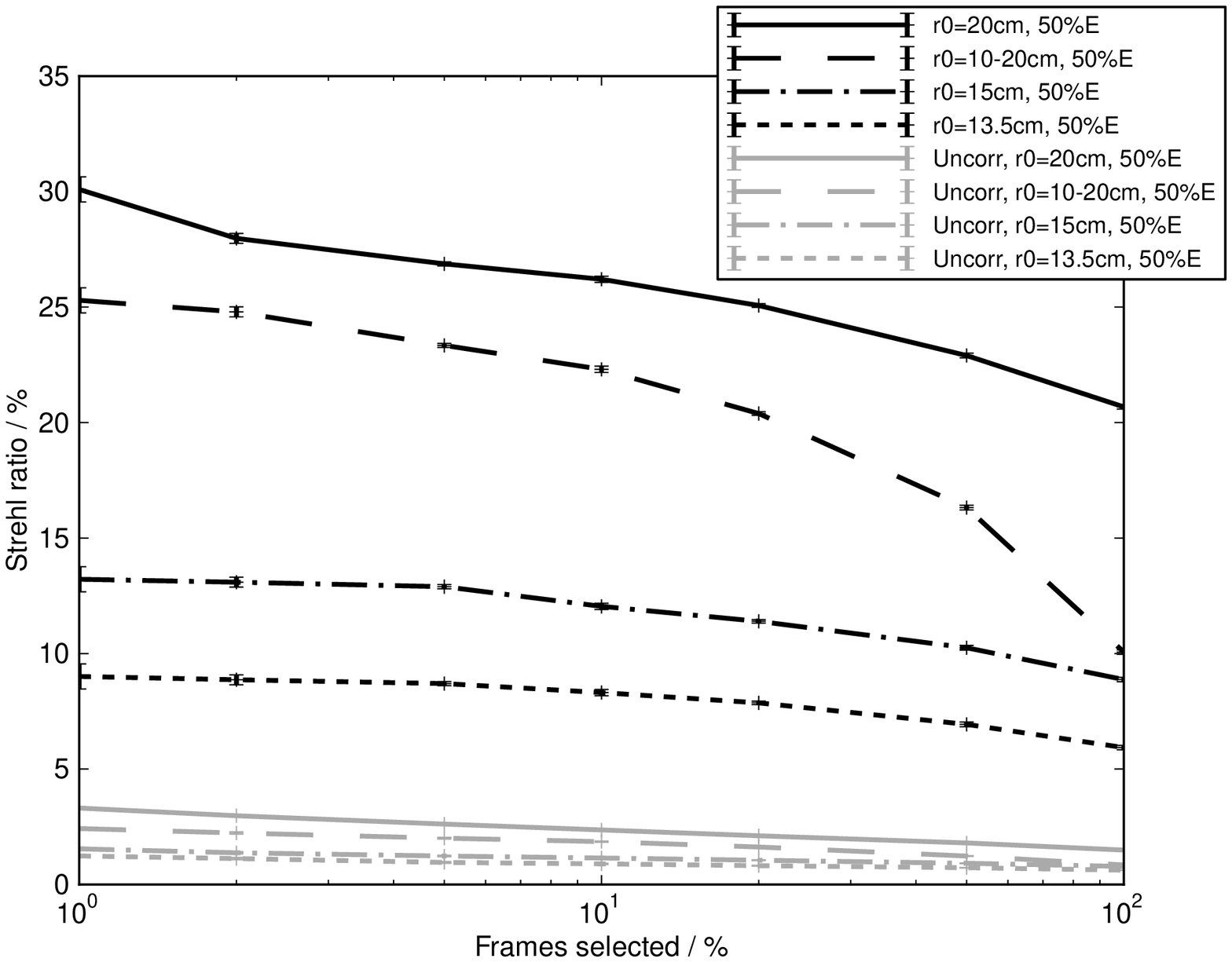}
\caption{A figure showing lucky image quality as a function of
  fraction of images selected for different atmospheric Fried's
  parameters as given in the legend, with the diameter encircling 50\%
  \psf energy used for image selection.  The AO corrected and
  non-corrected are shown, with ``Uncorr'' representing the
  un-corrected cases.}
\label{fig:r0d50}
\end{figure}

Lucky image quality when using a Strehl ratio criteria for image
selection also depends (as expected) on $r_0$.
Fig.~\ref{fig:r0strehl} shows a comparison of different atmospheric
conditions when lucky image Strehl ratio is used for image selection,
and again shows image quality falling more rapidly with selection
fraction when seeing is variable.  I-band Strehl ratios of 35\% have
been achieved with mean 15~cm $r_0$.  Using the \psf
diameter encircling 50\% energy as selection criteria leads to similar
trends, though with lower final Strehl ratio, as shown in
Fig.~\ref{fig:r0d50}.  

\subsection{Performance with sodium laser guide stars}
\begin{figure}
\includegraphics[width=\linewidth]{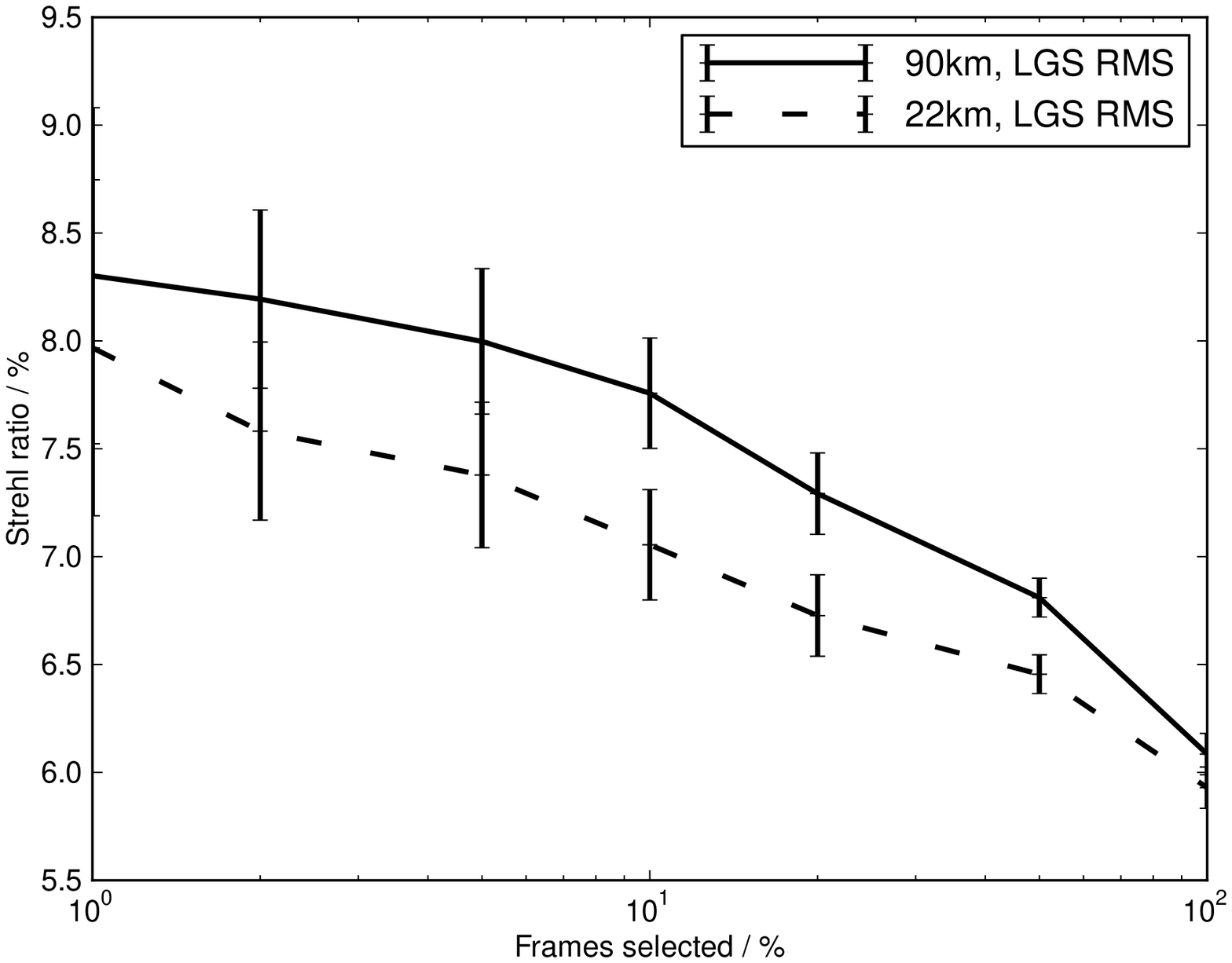}
\caption{A figure showing lucky image quality as a function of
  fraction of images selected comparing a Rayleigh (22km) and sodium
  (90km) laser beacon.  The long exposure Strehl ratio is about 1.5\%
  (LGS AO with no tip-tilt corrected).}
\label{fig:90km}
\end{figure}

Although we will be performing on-sky demonstration of this technique
using Rayleigh laser guide stars, we have also investigated
performance when using sodium laser guide stars, using a sodium layer
profile centred at 90~km above the telescope.  As shown in
Fig.~\ref{fig:asterism}, increased \lgs height results in increased
sampling of the atmospheric turbulence, thus leading to better
wavefront estimation and \ao correction.  Fig.~\ref{fig:90km} shows
lucky image performance comparing the 22~km and 90~km guide stars.  It
can be see that improved image quality is obtained when using 90~km
guide stars.  It should be noted that on the 4.2~m telescope simulated
here, there is significant pupil divergence at the highest atmospheric
layers even when using sodium \lgss, and hence a significant volume of
turbulence remains unsampled.  For larger telescope pupil diameters,
the atmospheric sampling would be more complete, and better performance
would be expected.

\subsection{Considerations for larger telescopes}
Although we have shown that the multiple \lgs-assisted lucky imaging
technique works well on 4~m class telescopes, it is important to
consider larger telescopes, including the forthcoming \elts.  The
increased diameter of these telescopes (over the 4.2~m case considered
here) will lead to more complete atmospheric sampling with the \lgss
(due to increase cone-effect diameter), and hence improved \ao
correction.  However, assuming that an \ao system has constant pitch
sub-apertures (typically 50-60~cm for current and proposed
multi-purpose \ao systems), then as telescope size increases, the
probability of \ao corrected residual wavefront phase perturbations
remaining small over the whole telescope aperture decreases.  Although
the \ao system will remove low and mid-order wavefront perturbations
(excluding tip-tilt), there is increasing probability that high order
perturbations, on scales less than a sub-aperture pitch, will remain
within some parts of the pupil.  Two techniques can be used to
mitigate this effect: Either \ao system order can be increased
(i.e.\ sub-aperture and \dm actuator pitch reduced) to reduce the area
of pupil over which significant unlucky turbulence remains.
Alternatively, an adaptive apodisation technique can be used, first
detecting areas of the pupil with significant residual wavefront
perturbation, and then selectively blocking these areas.  However,
both of these solutions introduce extra system complexity.

It should also be noted that depending on the required integrated
image performance metric, small areas of residual wavefront
perturbation on a large telescope pupil may not lead to a significant
degradation of image quality, and therefore, the fall in probability
with telescope diameter of obtaining a lucky image may be slow.
Additionally, the use of \lgs-only \mcao systems could further improve
the field of view over which lucky images can be obtained, though care
would be required to handle undetected ``breathing'' modes of these
systems.  

The full extension of the proposed \lgs-only \ao assisted lucky
imaging technique to telescopes with 8-40~m diameters requires further
study, which we do not seek to address here.

\subsection{Off-axis performance}
\begin{figure}
\includegraphics[width=\linewidth]{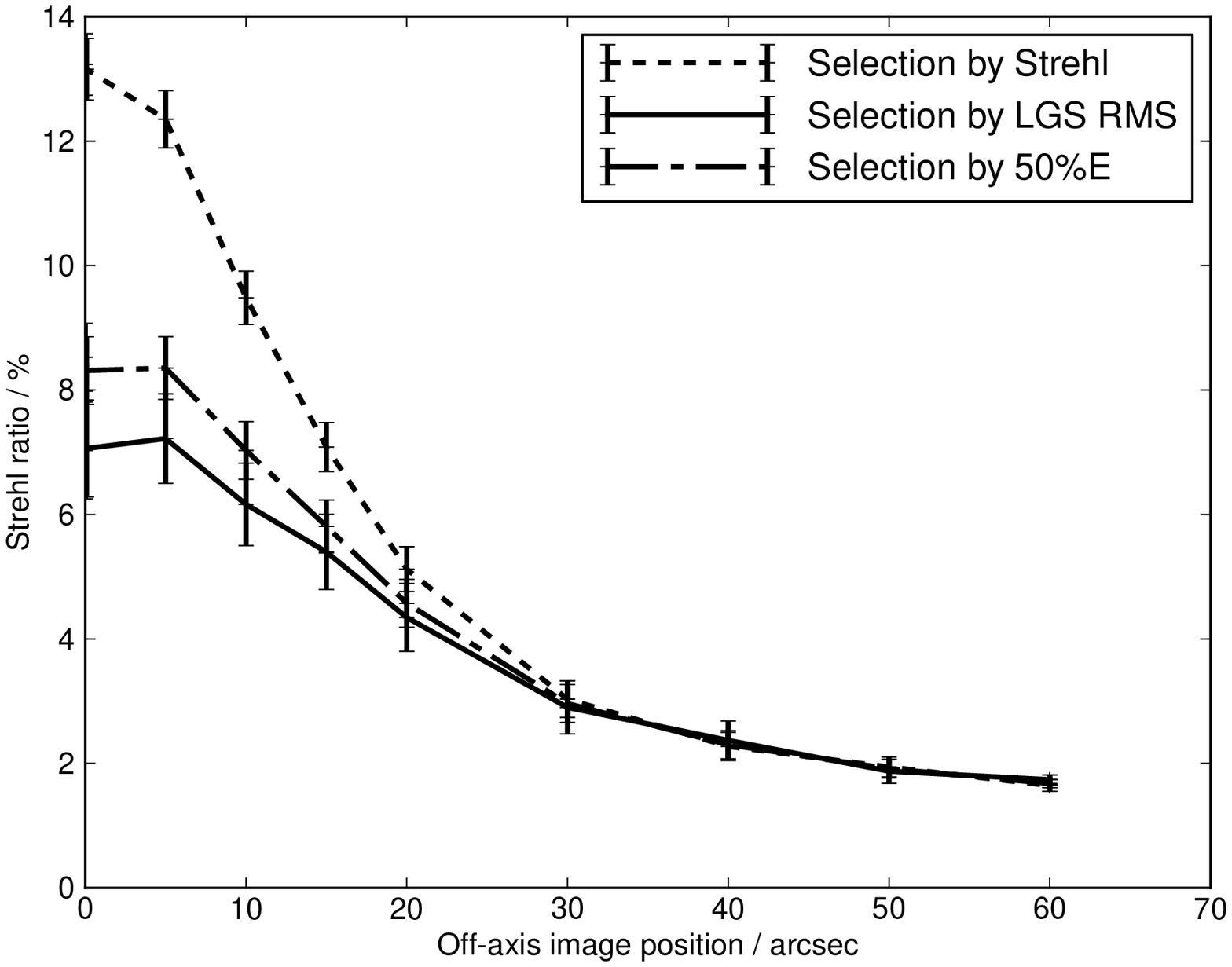}
\caption{A figure showing off-axis performance of \lgs assisted lucky
  imaging, with image selection and centring performed on-axis.}
\label{fig:onaxis}
\end{figure}

We now investigate the case where a central on-axis star is used for
image centring and selection of off-axis targets (which can then be
extremely faint).  Fig.~\ref{fig:onaxis} shows performance as a
function of off-axis distance (i.e.\ distance between the centring and
selection guide star, and the science target).  In the case where
image selection is performed using the \lgs rms wavefront error, this
selection is done along the line of sight of the target, since no flux
is required from the target.  However, image centring is still
performed using information from the on-axis guide star.  This figure
shows that performance falls most rapidly for selection based on
Strehl ratio, and that other selection methods offer more uniform
(though lower) performance across a field of view.

\section{On-sky testing of LGS assisted lucky imaging}
The CANARY instrument \citep{canaryshort} is an on-sky \ao technology
demonstrator, which has successfully demonstrated several different
\ao modes, including \moao and tomographic \glao, both with and
without \lgss.  CANARY has four Rayleigh \lgss, which are currently
range-gated at a height of 22~km.  It also has an \ao corrected
visitor instrument port, which has previously been used for a novel
spectroscopic instrument \citep{photonicLantern}, and provides an ideal test-bed
for developing \lgs assisted lucky imaging techniques.

We intend to use this visitor instrument port with a low noise \emccd
to capture lucky images at high frame rate, synchronised with \ao
telemetry data.  We will operate the lucky imaging camera using the
same real-time control system that is used by CANARY
\citep{basden9,basden11}, thus ensuring compatible data formats and
time-stamping.  CANARY will be operated in several \ao modes during
these observations, including \scao, \glao and \moao, with \ngs-only,
\ngs+\lgs and \lgs-only modes, and we will record lucky images for
both bright and faint targets so as to push lucky imaging towards
greater sky-coverage.  An on-axis truth \wfs will be used to record
on-axis wavefront phase for comparison with tomographically computed
measurements.  This wealth of combined \ao and lucky image data will
then undergo extensive post-processing to demonstrate the feasibility
of full-sky \lgs \ao-assisted lucky imaging.

\section{Conclusion}
We have introduced the concept of tomographic \lgs assisted lucky
imaging, using \ao techniques to extend the use of lucky imaging for
larger telescopes.  We use multiple \lgss to tomographically
reconstruct wavefront phase distortions due to atmospheric turbulence,
and apply a correction using a \dm, though without performing any
tip-tilt component correction.  Instantaneous images will therefore
have an improved \psf, but wander about.  Lucky imaging techniques are
then used to select the best images, and re-centre them, building up a
high resolution image.  The \ao correction does not require any
natural guide stars, and thus full sky-coverage is achieved.

We have investigated the use of wavefront phase as a lucky image
selection criteria, and find that it can offer improvements in image
selection for cases when signal-to-noise ratio is low in the lucky
images, though in the high signal-to-noise regime, conventional image
selection criteria provide better selection.  When field tip-tilt can
be determined, this method enables lucky imaging on very faint
sources.  

We find that there is some advantage in reducing \wfs pitch relative
to \dm pitch which can offer improved performance in cases when \lgs
flux is not limited, due to better estimation of wavefront phase.  We
have performed modelling with constant and variable seeing, and as
expected find that lucky imaging benefits periods of good seeing
within a variable seeing case, particularly when the fraction of
images selected is low.  We also find that using sodium \lgss can offer
improved performance over Rayleigh \lgss due to increased sampling of
atmospheric turbulence.

We have demonstrated that the concept of tomographic \lgs assisted
lucky imaging has potential to yield near-diffraction limited visible
optical (I-band) images on 4~m class telescopes.  Strehl ratios of up
to 35\% have been achieved in simulations with mean seeing of about
0.7~arcsec (with a mean $r_0$ of 15~cm, varying sinusoidally from 10
to 20~cm).  There is unlimited sky-coverage for the \ao, since this
concept relies only on \lgs.  Our simulations, which have included
photon shot noise and detector readout noise, have shown that when
using a crude image centring and selection criteria, stars as faint as
I-mag 18 can be used, depending on image quality metric.  On larger
telescopes, fainter targets can be reached, since \ao corrected,
rather than seeing-limited, instantaneous \psfs are used, which
concentrates the photon flux into a smaller diameter core as telescope
diameter increases, yielding improved signal levels.  Increasing
system bandpass, to increase photon flux onto the lucky imaging camera,
will also increase the achievable limiting magnitude, though
achromatic wide field of view optical design is nontrivial.  This
technique is suitable for use with all forms of tomographic \ao
systems, including \moao, \mcao, \glao and \ltao.

\section*{Acknowledgements}
This work is funded by the UK Science and Technology Facilities
Council, grant ST/K003569/1.  The author would like to thank Richard
Myers and Craig Mackay for insightful comments.

\bibliographystyle{mn2e}

\bibliography{mybib}
\bsp

\end{document}